# Consistency of GPS and strong-motion records: case study of Mw9.0 Tohoku-Oki 2011 earthquake

Panos Psimoulis[1,2*], Nicolas Houlié[2,3], Michael Meindl[2], and Markus Rothacher[2]

[1]*Nottingham Geospatial Institute, The University of Nottingham, Nottingham NG7 2TU, UK*
[2]*Geodesy and Geodynamics Lab., Geodesy and Photogrammetry Institute, ETH Zurich, Zurich 8093, Switzerland*
[3]*Seismology and Geodynamics, Institute of Geophysics, ETH Zurich, Zurich 8092, Switzerland*

**Abstract.** GPS and strong-motion sensors are broadly used for the monitoring of structural health and Earth surface motions, focusing on response of structures, earthquake characterization and rupture modeling. Most studies have shown differences between the two systems at very long periods (e.g. >100sec). The aim of this study is the assessment of the compatibility of GPS and strong-motion records by comparing the consistency in the frequency domain and by comparing their respective displacement waveforms for several frequency bands. For this purpose, GPS and strong-motion records of 23 collocated sites of the Mw9.0 Tohoku 2011 earthquake were used to show that the consistency between the two datasets depends on the frequency of the excitation, the direction of the excitation signal and the distance from the excitation source.

## 1. Introduction

The accurate estimation of the transient GPS antenna motion (i.e. error<1cm) for the monitoring of high-rate (i.e. up to 4Hz) oscillations, followed by the introduction of high-rate GNSS measurement noise modeling (Moschas and Stiros, 2013a), the improved processing modes and algorithms (Psimoulis and Stiros, 2012; Wang and Rothacher, 2013) and the time-stamping of observations (Psimoulis and Stiros, 2013) led to its application in structural health monitoring (Psimoulis et al., 2008; Psimoulis and Stiros, 2008). Mainly, in the field of structure monitoring, high-rate GPS is used as complementary of other motion sensors, such as accelerometers (Roberts et al., 2004; Chan et al., 2006; Meng et al., 2007; Chatzi and Fuggini, 2012; Moschas and Stiros, 2013b; Yi et al., 2013) or Robotic Total Station (Moschas et al., 2013). The accurate estimation of the transient GPS antenna motion is important not only for the understanding of structures deformations (Panagiotakos and Fardis, 1999; Penucci et al,. 2009; Sadan et al., 2013), but also for the detection of tectonics motions (Chousianitis et al., 2013), and the characterization of deformation transients during or after the earthquake (Larson et al., 2003; Miyazaki et al., 2004; Blewitt et al., 2006; Feng et al., 2008; Ganas et al., 2009; Houlié et al., 2011; Yue and Lay, 2011; Wright et al., 2012; Houlié et al., 2014).

*Corresponding author, Lecturer in Geospatial Eng., E-mail: Panagiotis.Psimoulis@nottingham.ac.uk

Until recently, the estimation of the displacement of the structure response and the corresponding excitation was based on accelerometers records (Wang et al., 2003, Feng, 2009). Such displacements time-series are, however affected by the seismic sensors saturation, significantly dominant for large excitation or large earthquakes (i.e. Sagiya et al., 2011) and characterized by systematic drift (Boore et al., 2003; Wang et al., 2003; Wang et al., 2007; Cauzzi and Clinton, 2013), which is increasing significantly with the duration of the acceleration record (Stiros, 2008).

The developed Precise Point Positioning (PPP) processing mode of GNSS data, can produce displacement waveform of cm-accuracy level based on standalone GNSS station (Ge et al., 2008; Geng et al., 2011; Joakinen et al., 2013), improving the conditions of utilization of GNSS as real-time monitoring system. The quality of time-series became sufficient to use GPS time-series to supplement earthquake early-warning systems (Crowell et al., 2009; Geng et al., 2013) and monitor remote structures (e.g. wind turbines; Giti and Lee, 2013).

Still, the consistency of PPP GPS with strong-motion time-series have never been fully tested for oscillations in a wide range of frequencies (from 0.001 up to 1Hz) such those generated by a seismic source. Such effort aims at i) assessing the displacement drift characteristic due to the double integration of acceleration records (Wang et al., 2003), ii) estimating the error propagation law explaining partly the accumulated error leading to the drift (Stiros, 2008), iii) quantifying the clipping of the strong-motion sensors (Clinton and Heaton, 2003) and finally iv) combining GPS and accelerograph to obtain single broadband displacement time series (Wang et al., 2007).

In this study, we use the data collected by three dense geophysical monitoring networks: the two strong motion networks of K-NET and KiK-net, and the GPS network of Japan during the Mw9.0 Tohoku-Oki 2011 earthquake. We focus on records recorded by closely located GPS and strong-motion sites. Data were evaluated in the frequency domain, based on the coherence analysis and by comparing the displacement waveforms for several frequency bands and for various distances to the from the epicentre.

## 2. Data

### *2.1 GPS data*

The Geospatial Information Authority of Japan (GSI) operates over 1200 continuously observing GPS receivers, known as GPS Earth Observation Network (GEONET), covering the Japanese land area with an average distance of about 20 km between neighboring points (Sagiya, 2004). The Mw9.0 earthquake off the Pacific coast of Tohoku on March 11, 2011 was well recorded by GEONET.

GPS records from 847 GEONET stations (Fig.1) of 15-hour duration and 1Hz sampling rate, covering the period of the earthquake, were available and processed using the scientific Bernese GPS Software 5.0 (Dach et al., 2007). The data were post-processed in a Precise Point Positioning (PPP) mode using a-priori information of highest quality from the Center for Orbit Determination in Europe (Dach et al. 2009), resulting in displacement time series in North, East, and vertical components of 1Hz sampling rate with a formal a-posteriori accuracy of about 1cm in the horizontal and 2cm in the vertical component.

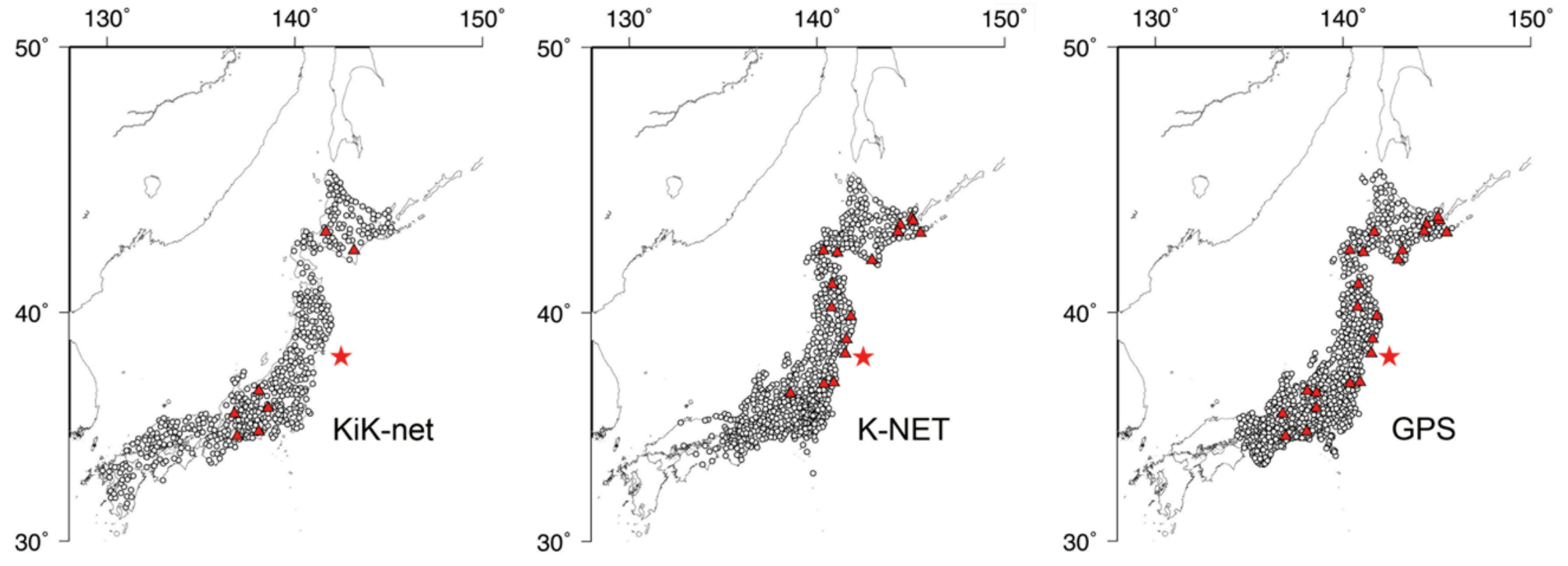

Fig. 1 The 23 collocated sites of the KiK-net/K-NET strong-motion networks with the GEONET GPS network (red triangles), as part of the available data of the three networks (white circles). The earthquake epicentre is marked as red star.

### *2.2 Strong motion data*

The strong-motion networks of Japan, K-NET (Kyoshin) and KiK-net (KIBAN Kyoshin) are composed of 1034 and 660 seismometers, respectively. The K-NET is deployed mostly on thick sedimentary ground, while the KiK-net is located on rock or thin sedimentary sites, as it is designed to capture high-accuracy seismic observation. Furthermore the triggering threshold of the KiK-net sensors is 0.2 cm/s$^2$, while that of K-NET sensors is 2 cm/s$^2$ (Aoi et al., 2004). K-NET is consisted of only surface stations, the KiK-net is consisted of two seismometers, installed at surface ground and at the bottom of boreholes of 100-200m depth.

The earthquake of Mw9.0 Tohoku 2011 was recorded successfully by the two strong-motion seismic networks and the raw data from 700 K-NET sites and 525 KiK-net sites were available (Fig.1). The sampling rates of records is 100Hz, their duration up to 300sec. The acceleration of the each site derived by amplifying the raw data according to a scale factor and the corresponding record time, initially given in UTC time, was corrected (i.e. for 15 leap seconds) and transformed into GPS time.

### *2.3 Very close spaced GPS – strong motion sites*

The evaluation of the consistency of the GPS and the strong-motion records is made using data collected by very close spaced sites of the GPS and strong-motion networks. At high frequency (> 1Hz), in order to complete a valid comparison, the seismic and geodetic sensors need to be as close as possible. GPS antennas are often not perfectly collocated (less than 1m) with seismic sensors to improve the quality of the seismic data and to maximize the number of GPS satellites potentially visible in the sky. GPS and strong-motion sites are assumed as collocated when the distance between the two instruments is less than 140m. Considering the minimum distance between the epicenter and the network (~200km for the GPS network), and the distance between the two sensors (i.e. <150m), we assume that the same seismic motion was recorded by both GPS and strong-motion sensor without significant time difference between the two sensors records.

Amongst the three networks, we identified 23 pairs of very close spaced sites of the GPS and strong-motion networks (16 K-NET and 7 KiK-net network; Table 1), without any malfunctions or delayed triggering (Fig. 1). The 23 collocated sites cover the entire range of the distance from the epicentre (50-700km) of the three networks. For simplicity from this point and on these 23 very close spaced sites will be referred as collocated sites from this point on.

### *2.4 Definition of the GPS and strong-motion time series*

The acceleration time series of the strong-motion was defined according to the start and end time of the corresponding record, including the transformation from UTC to GPS time. Regarding the GPS time series, they were defined according to the start and end time of the strong-motion record of the corresponding collocated site. The latter certified the common reference time and the compatibility of the time series of the GPS and strong-motion site.

Table 1: The collocated GPS and strong-motion sites of K-NET and KiK-net network with their location (latitude, longitude), distance from the epicentre and the distance between the two collocated sensors

| GPS | K-NET | KiK-net | latitude | longitude | Sensors distance (km) | Epicentre distance (m) |
|---|---|---|---|---|---|---|
| 5 | HKD080 | - | 43.50786 | 144.44901 | 604 | 44 |
| 41 | FKS011 | - | 37.09071 | 140.90252 | 186 | 97 |
| 115 | HKD066 | - | 43.66182 | 145.13143 | 639 | 59 |
| 122 | HKD083 | - | 43.23294 | 144.32503 | 572 | 19 |
| 140 | HKD151 | - | 42.49436 | 140.35420 | 496 | 24 |
| 144 | HKD110 | - | 42.13094 | 142.93539 | 428 | 38 |
| 164 | IWT019 | - | 39.84918 | 141.80385 | 179 | 39 |
| 183 | AKT006 | - | 40.21544 | 140.78733 | 253 | 26 |
| 519 | HKD072 | - | 43.19517 | 145.52050 | 605 | 52 |
| 535 | AOM027 | - | 41.14555 | 140.82198 | 343 | 28 |
| 792 | HKD131 | - | 42.42055 | 141.08106 | 471 | 67 |
| 793 | - | TKCH08 | 42.48641 | 143.15197 | 470 | 10 |
| 864 | HKD065 | - | 43.79403 | 145.05686 | 650 | 38 |
| 877 | - | IKRH02 | 43.22090 | 141.65202 | 550 | 60 |
| 983 | - | NGNH28 | 36.70665 | 138.09673 | 416 | 98 |
| 991 | - | GIFH23 | 35.72349 | 136.78462 | 572 | 7 |
| 998 | - | AICH21 | 34.74005 | 136.93848 | 625 | 6 |
| 550 | MYG011 | - | 38.30119 | 141.50069 | 75 | 131 |
| 172 | MYG001 | - | 38.90286 | 141.57259 | 96 | 116 |
| 591 | GNM004 | - | 36.61632 | 138.59145 | 382 | 89 |
| 613 | - | NGNH19 | 35.97415 | 138.58303 | 423 | 129 |
| 945 | FKS015 | - | 37.02358 | 140.37627 | 225 | 123 |
| 3088 | - | SZOH31 | 34.93980 | 138.07486 | 534 | 147 |

## 3. Methodology

### *3.1 Coherence analysis*

The coherence analysis is a broadly used technique for the cross-spectral density of two data-set in the frequency domain (Kim and Stewart, 2003; Mikami and Stewart, 2008; Moschas and Stiros, 2011) and it was used to assess the consistency of collocated GPS and strong-motion sites. The magnitude squared coherence was used, which is based on Welch's average modified periodogram, provided by Matlab. The magnitude squared coherence ranges between 0 and 1, considering that for coherence greater than 0.8 the two time series can be assumed strongly correlated in the frequency domain, while deviating coherence below 0.8 indicates significant noise spectrum (Kim and Stewart, 2003).

We apply the coherence analysis for the common frequency range (0-0.5Hz) of the 1Hz GPS and 100Hz strong-motion records. However, before comparison, strong-motion data were decimated from 100 to 1Hz resulting finally to compatible GPS and strong-motion records of 1Hz sampling-rate and 300sec duration. For the coherence analysis was used Hamming window of 100 samples, with overlapping of 90 seconds in order to cover broad range of frequencies (i.e. 0.005-0.5Hz).

### *3.2 Displacement waveforms comparison*

The frequency bands for the assessment of the GPS and strong motion displacement time series were formed, based on an octave-wide distribution (Clinton and Heaton, 2003, Cauzzi and Clinton), the limitations of the sampling rate and the duration of the records. Specifically, the upper frequency limit was defined 0.4Hz, smaller than the Nyquist frequency and not close to the relative high frequency (i.e. 0.5Hz), where the GPS noise level is expected to be increased. Following the octave-wide distribution and having the 0.4Hz as upper limit, there were formed 6 frequency bands for the assessment: i) 0.2-0.4Hz, ii) 0.1-0.2Hz, iii) 0.05-0.1Hz, iv) 0.025-0.05Hz, v) 0.0125-0.025Hz, and vi) 0.0061-0.0125Hz. The lowest frequency limit was 0.0061Hz, in order to avoid the low-frequencies (<0.005Hz) of the displacement time series of the strong-motion sensors which would suffer strongly by drifting (Wang et al., 2003), especially in the case of the examined strong-motion records due to their large record duration (e.g. 300 sec).

The second step was the band-pass filtering, which was done by using Chebyshev filter of $8^{th}$ order, which have more abrupt cut-off frequency than Butterworth filter (Boore and Bommer, 2005), without distorting significantly the data (Moschas and Stiros, 2013b). The filter was applied twice, in forward and reverse direction, in order to limit potential phase shift of the derived time series. The strong-motion and the GPS displacement time series were filtered for each one of the examined frequency bands, resulting to six new displacement time series of GPS and strong-motion for every component of each collocated site. The application of the band-pass filtering to the strong-motion displacement time series for the examined frequency bands led to the limitation of the low-frequency drifting effect.

The next step was the cross-correlation analysis of the corresponding GPS and strong-motion displacement time series of each component (NS.EW,UP) and each frequency band of the collocated sites, for the detection of possible time lags. The latter can be due to the distance between the GPS and strong-motion sites or even clock drift of the strong-motion sensor, which cannot be regularly corrected (Moschas and Stiros, 2012).

Finally, based on the estimated time lag for each collocated displacement time series, the corresponding time series were shifted and compared by computing their relative difference, known as residuals. The obtained residuals were assessed by computing the corresponding standard deviation and their maximum value.

## 4. Results

### *4.1 Coherence Analysis*

We show in Fig. 2 the plots of coherence of six representative collocated sites, covering the distance range of 0-700km from the epicenter. The collocated sites (GPS:550 with K-NET:MYG011), which are close (<200km) to the epicentre, appears high coherence (>0.8) on the vertical component for the frequency range of 0.05-0.25Hz, whereas the corresponding frequency range of high coherence (i.e. >0.8) for the horizontal components is rather limited (0.05-0.1Hz). For collocated sites of moderate distance from the epicentre (i.e. 350km; GPS:183 with K-NET:AKT006), high coherence is observed both in horizontal and vertical components for frequencies ranging between ~0.025 and 0.3Hz, while far from the epicentre (i.e. 640km; GPS115 with K-NET:HKD066), the range frequencies of high coherence of the horizontal components remains the same (~0.025-0.3Hz) and that of the vertical component is limited significantly (0.03-0.1Hz).

Thus, the coherence analysis of the GPS and strong-motion data of the collocated sites reveals that the frequency range of high coherence (>0.8) depends on the component and the location of the site, with respect the earthquake epicentre, which indicates that the consistency of the GPS and strong-motion records of the collocated sites, is relative with the directivity of the propagation of the seismic waves.

### *4.2 Displacement waveforms comparison*

The displacement waveform analysis was applied for each one of the frequency bands and for each component of the 23 collocated GPS and strong-motion sites. Initially, the time lags between the GPS and strong-motion corresponding displacement time series were estimated, ranging between -2 and 3 seconds. The plot of the time lags of the three components of all frequency bands versus the distance from the epicentre shows that the time lags are independent of the distance from the epicentre and less dispersion appears at the time lags of the vertical component relatively to that of the horizontal components (Fig. 3left). The larger time lag for each collocated site appears generally at the highest frequency band for the horizontal components, while for the upward component this appears at the lowest frequency band. Furthermore, the time lags seem to be independent of the distance between the collocated sensors (Fig. 3right), indicating that the latter did not affect the collocation conditions of the two sensors.

Based on the computed time lags of the collocated sites, the corresponding displacement time series were shifted for the "synchronisation" and the corresponding residuals were computed. In Fig. 4 are plotted the representative GPS and strong-motion displacement time series of the EW component and the corresponding residuals derived from the filtering for two frequency bands

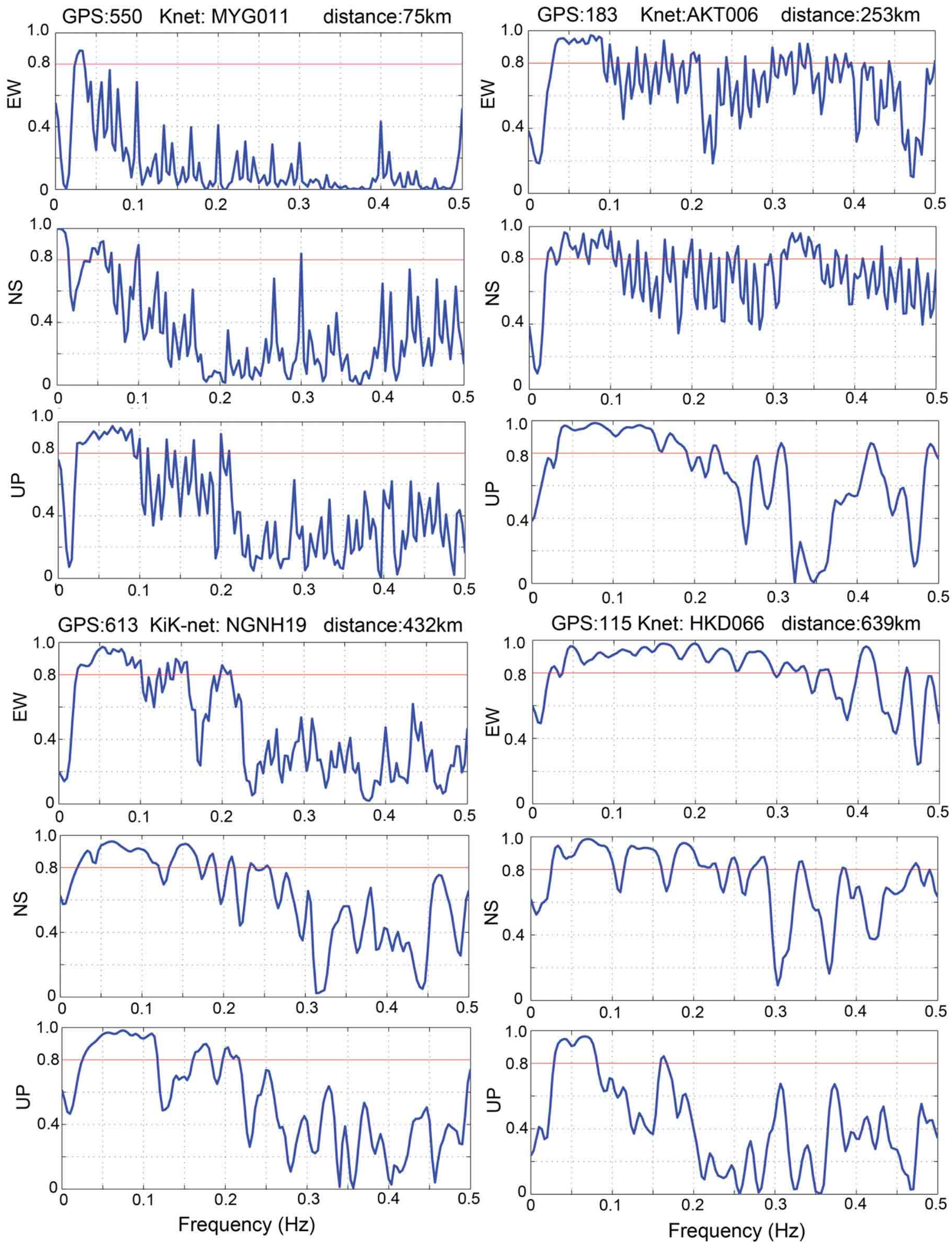

Fig. 2 Plots of the coherence analysis of 6 representative GPS and strong-motion collocated sites with distance from the epicentre from 75 to 640km. The horizontal line indicates the limit of 0.8 for the assessment of coherence.

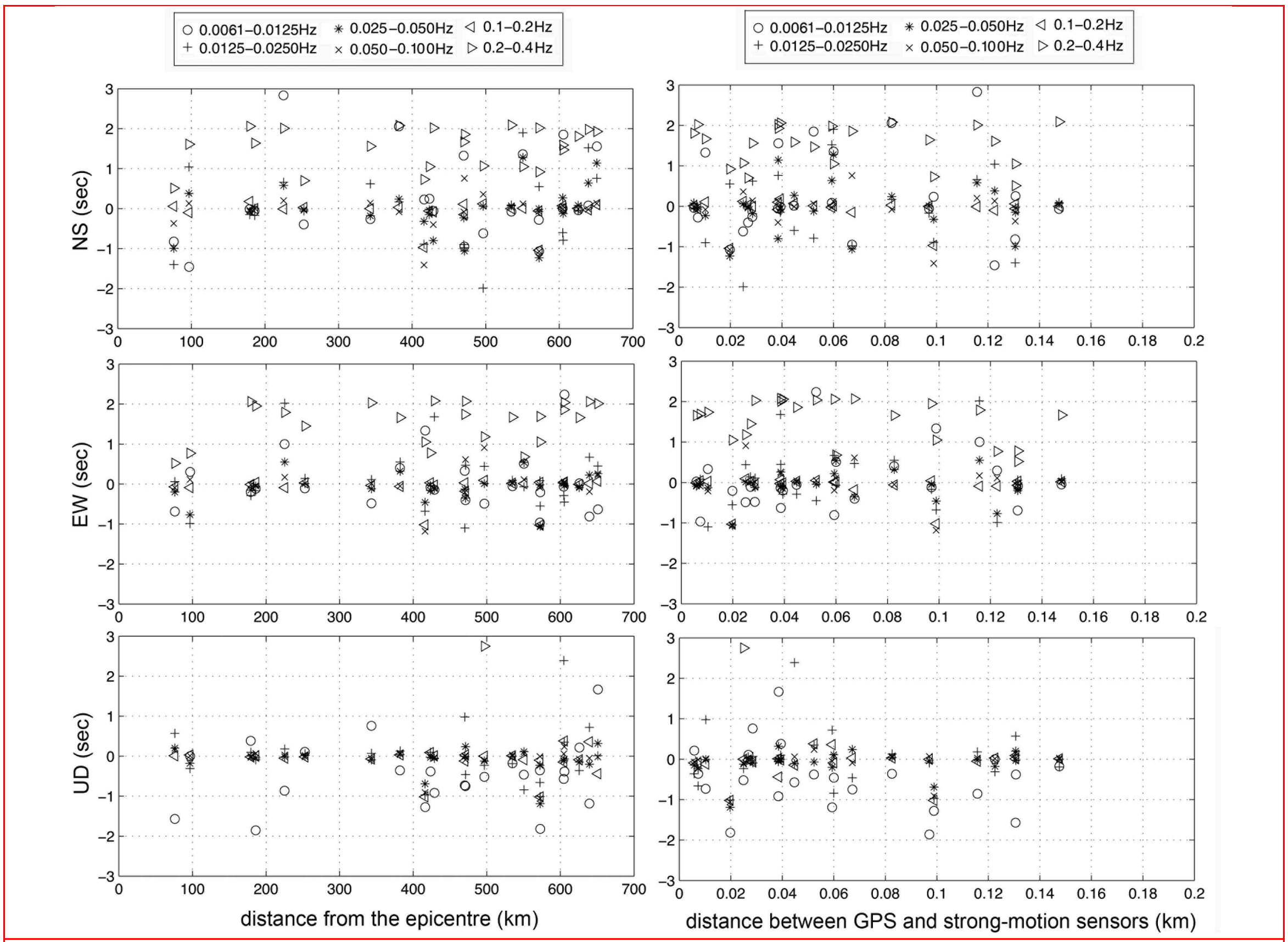

Fig. 3 The computed time lag between the GPS and strong-motion records of the collocated sites versus the corresponding distance from the epicentre (left) and the corresponding distance between the two sensors (right).

(0.0061-0.0125Hz and 0.225-0.5Hz), for three collocated sites (i.e. GPS:550 with KNET:MYG011, GPS:164 with K-NET: IWT019 and GPS:613 with KiK-net:NGNH19). The first impression of the GPS and strong-motion displacement time series reveal rather small relative amplitude difference (i.e. ~6-7cm for the GPS550 and K-NET MYG011), which finally proved to be significantly larger according to the corresponding computed residuals. The amplitude of the residuals decreases with the increase of the frequency band and the distance from the epicentre. This is made clearer by computing the standard deviation (Fig. 5) and the maximum value of the residuals (Fig. 6left). The standard deviations range mainly between a few sub-millimeters up to 10cm, while the maximum estimated residuals range between 1mm and 30cm. By excluding the collocated sites very close to the epicentre (<100km), the computed standard deviation and maximum residuals ranges are limited up to 4-5cm. In both plots there is a pattern of decrease of the standard deviation and the maximum residuals with the distance from the epicentre, while the computed values of the upward component are smaller than these of the horizontal, mainly for distance close to the epicentre (i.e. <100-150km).

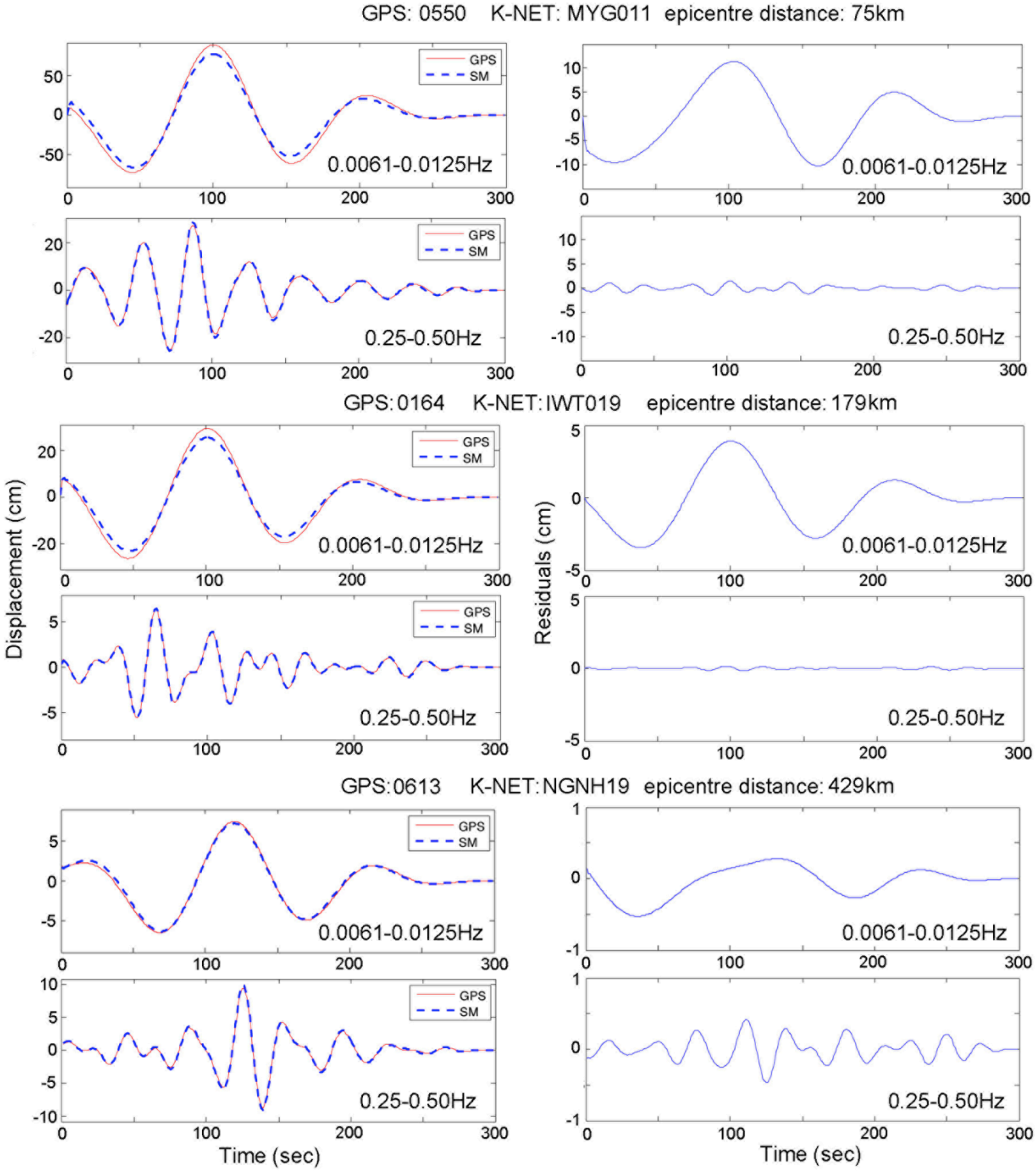

Fig. 4 Displacement of three collocated GPS and strong-motion sites for two frequency bands (0.0061-0.0125Hz and 0.25-0.50Hz) for three distances from the epicentre and the corresponding computed residuals.

Furthermore, the largest maximum residuals and standard deviations correspond to the lowest frequency band for distance smaller than 400km from the epicentre. For large epicentre distance (>400km) the distribution of the frequency bands in the range of the estimated standard deviation and maximum residuals seems rather random.

However, the normalised maximum residuals, based on the maximum GPS displacement of the corresponding component, are generally increasing with the distance from the epicentre, ranging between 0.002 to 0.25 (Fig. 6right). Again the largest normalised residual is observed for the lowest frequency band for the distances less than 400km from the epicentre, while for sites far from the epicentre (>500km), the correspondence is rather random.

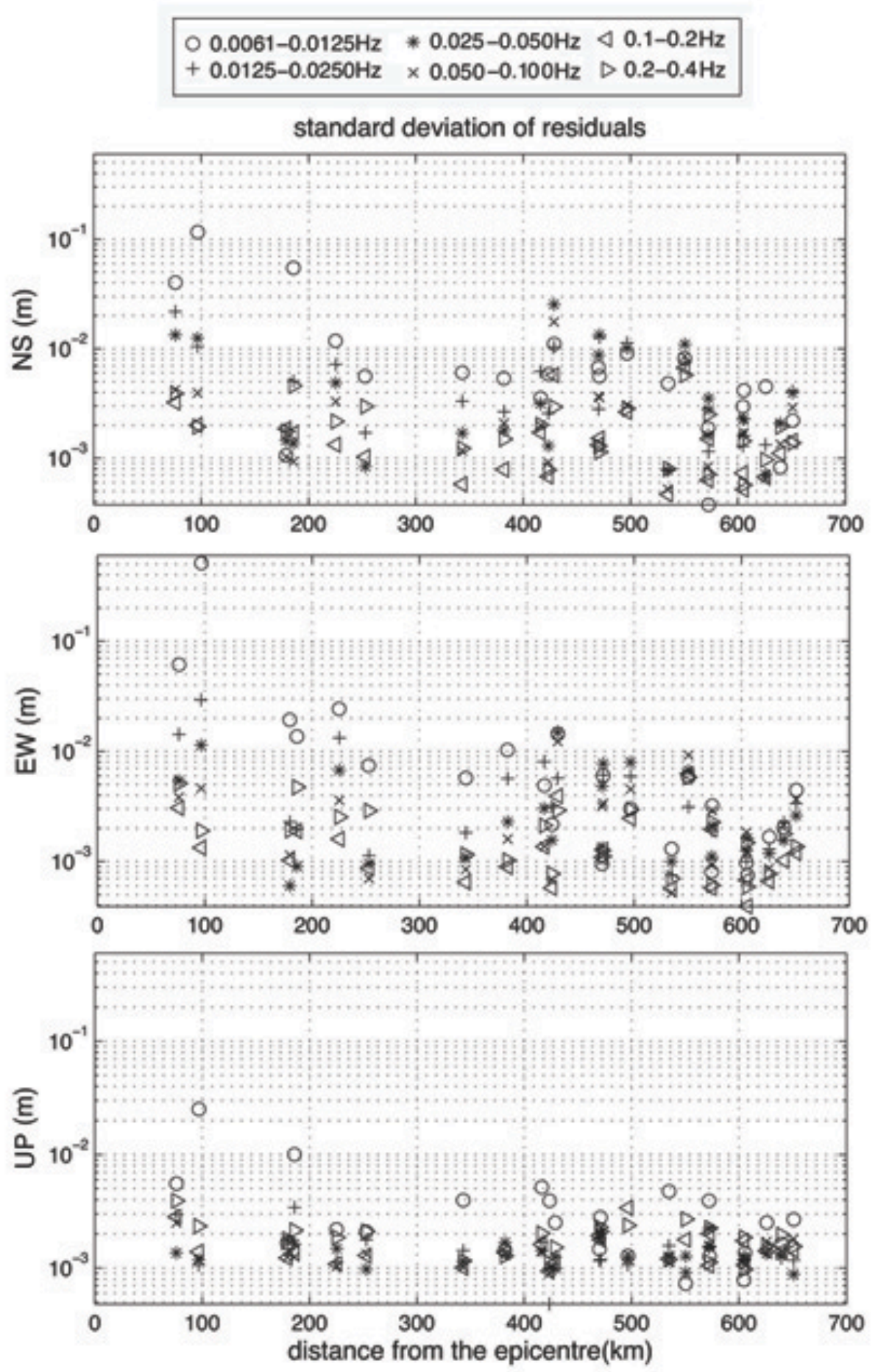

Fig. 5 Standard deviation of the residuals for every component (NS, EW, UP) of the collocated sites versus the corresponding distance from the epicentre.

The residuals were further analysed, using the Discrete Fourier Transform (DFT), which generally reflected the distribution of the frequency bands in the range of the computed standard deviations. In Fig. 7 we present the spectra of residuals of the six frequency bands for two representative collocated sites (GPS:613 with KiK-net:NGNH19 and GPS:164 with K-NET:IWT019), where the frequency bands greater then >0.05Hz reveal weak signals (of magnitude of sub-mm) with quite random frequency peaks expressing mainly noise of the measurements or even caused by possible data distortion from the band-pass filtering. For the frequency bands greater than 0.05Hz, the revealed frequency peaks become more dominant especially for the sites relatively close to the epicentre (i.e. spectrum of EW of GPS:172 K-NET:MYG001 of frequency band of 0.0061-0.0125), appearing also consistency in some cases between the components of the site (i.e. EW-NS consistency of the detected frequencies for site GPS:613 K-NET:MYG001 for frequency band 0.0125-0.025Hz).

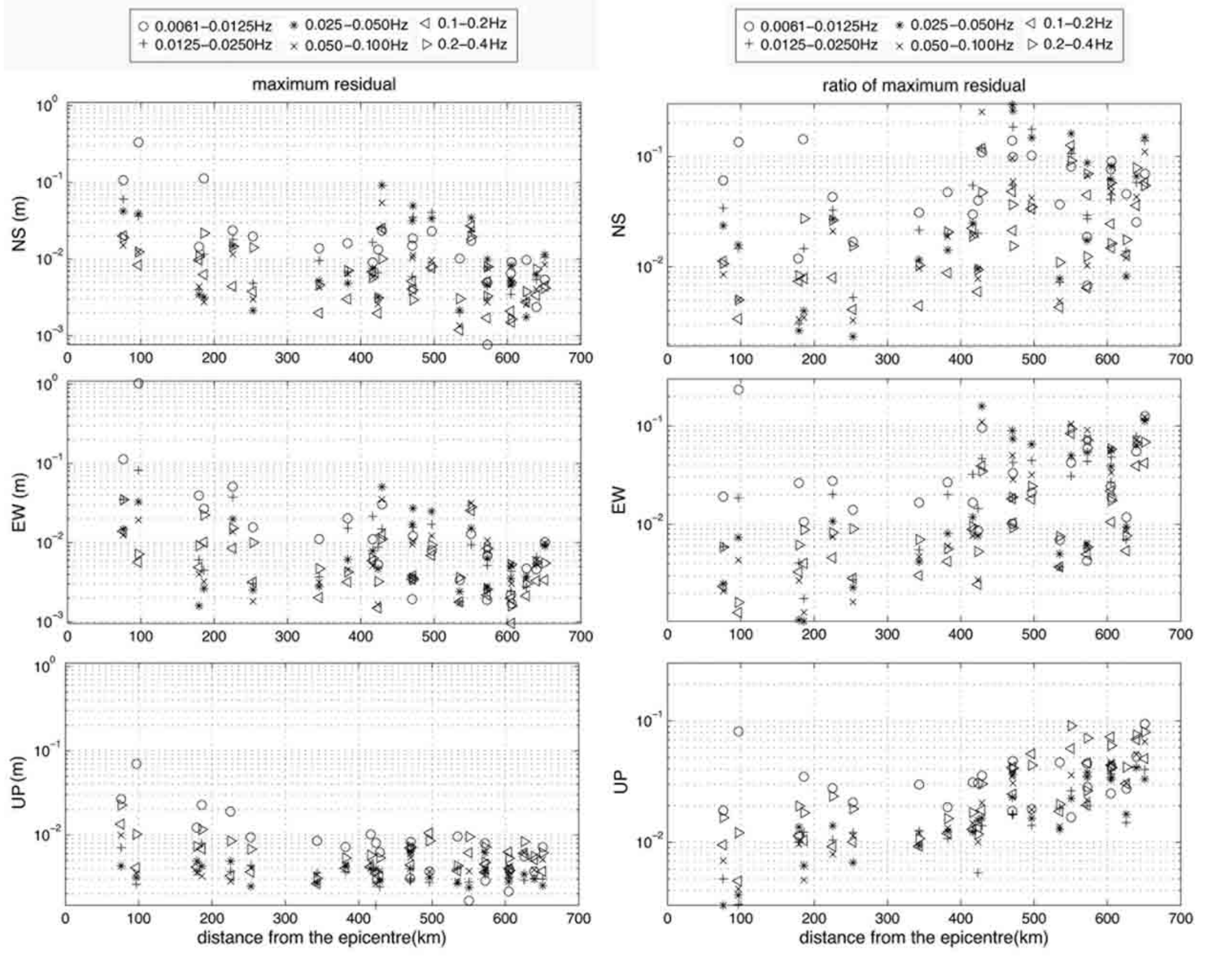

Fig. 6 Plot of the maximum residual for every component (NS, EW, UD) of each collocated site versus distance from the epicentre (left). Plot of the maximum residual normalized to the corresponding maximum displacement for every component of the collocated sites versus the distance from the epicentre (right).

## 5. Discussion

The assessment of the consistency of GPS and strong-motion records using the coherence analysis in the frequency domain and the displacement comparison in the time domain for several frequency bands showed that the consistency of GPS and strong-motion sensors depends on the distance from the epicentre. For sites close to the epicentre (<100-200km), higher consistency of GPS and strong-motion sensors appears more often on the vertical than the horizontal components, which reverses away from the epicentre (i.e. consistency for horizontals better than vertical component). Again, the reliability of the vertical components of motions appears to be better than the horizontal ones, likely because of the anisotropic distribution of GPS satellites in the sky (Houlié et al., 2011). The latter agrees with the directivity of the seismic wave propagation (Shearer, 1999), which is dominant on the vertical and horizontal components, close and far from the epicentre, respectively.

Furthermore, the coherence analysis showed that for a frequency range of 0.025-0.1Hz, we estimate a high coherence (>0.8) between the two signals whatever the component or the epicentre distances, indicating a frequency range of the best-expected consistent GPS and strong-motion records.

The large maximum residuals of the relatively close distant (<400km) collocated sites corresponding to band of 0.0061-0.0125 Hz can be attributed mainly to the phase shift between the GPS and strong-motion time series. The phase shift is not constant and it is probably due to random phase delays of the accelerometer, independently from the motion characteristics (Moschas and Stiros, 2012). This effect decreases with the increase of the frequency, due to the weak displacement signal of this frequency band.

However, for collocated sites located far from the epicentre, weak displacement signal in combination with GPS noise level may lead to poor correlation between the two families of signals, suggesting GPS is not the best tool to use to constrain displacement where the displacement is close to the formal error of PPP processing.

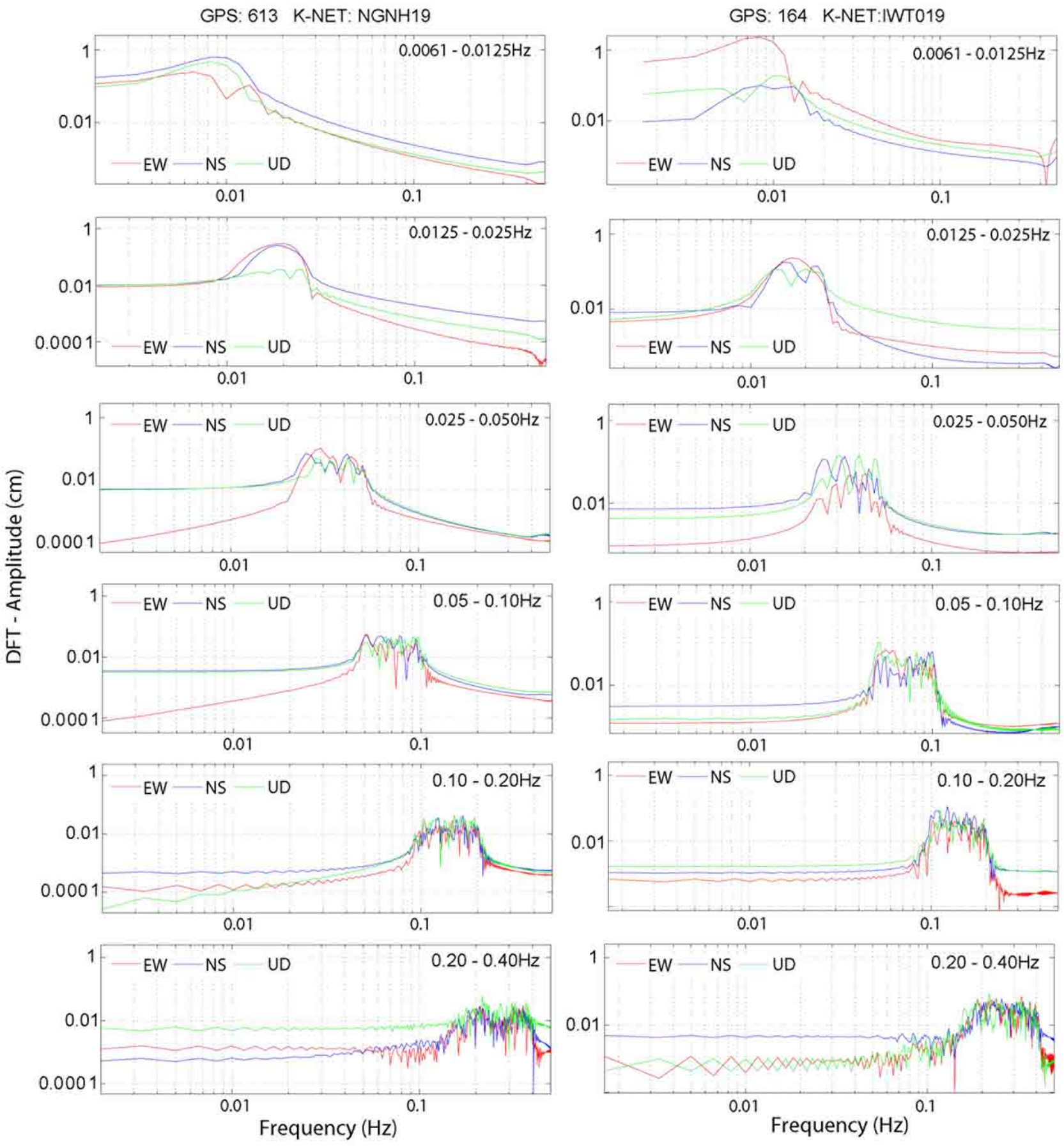

Fig. 7 The spectra of the residuals of the displacement of two representative collocated sites: GPS:613 with KiK-net:NGNH19 and GPS:164 with K-NET:IWT019.

To summarize, the ratio of the residuals of GPS and strong-motion displacement difference to the maximum corresponding GPS displacements, for all the frequency bands, increase with the distance from the epicentre, which is due to the attenuation of the seismic signal (Fig. 2), resulting to deviation corresponding up to 20% of the maximum corresponding displacement far from the epicentre (>400km). The high ratio of the residuals to maximum GPS displacement (up to 10%)for the sites (i.e. GPS550 with K-NET MYG011 and GPS179 with K-NET MYG001) very close to the epicentre (<100km) correspond to the NS component and apart from the phase shift it may also be attributed to clipping of the strong-motion sensor and/or tilting of the sensor contributing wrongly to the recording of the NS component. Generally, the GPS and the strong-motion displacement waveforms can calibrate and correct the one the other, by using the GPS for the low frequencies (<0.01Hz) and permanent displacement and the strong-motion for the relative high-frequencies (>0.3-0.4Hz), considering also the component and the relative location from the epicentre.

## 6. Implications for Structural Health Monitoring and Earthquake Engineering

Previous studies have shown the significance of the displacement waveforms for the design of the structure (Panagiotakos and Fardis, 1999; Adhikari et al., 2010; Penucci et al,. 2009; Sadan et al., 2013) and revealed the need of using geodetic techniques and especially GNSS technology. The combined use of GNSS and strong-motion sensors (accelerometers/seismometers) is not only for covering the drawback of the strong-motion sensors in the very low-frequencies (<0.01Hz; Wang et al., 2003). The consistency of GPS and strong-motion sensors in specific frequency range of motion gives the potential to overcome their malfunctions (e.g. rotation of the strong-motion sensors, errors in GPS measurement due to poor satellite constellation, etc.) and supplement each other by calibrating and the one the other their errors, considering also the location relatively to the earthquake epicentre. The calibration, already applied with geodetic measurements (Moschas et al., 2013), lead to the creation of "combo" stations including both GNSS and strong-motion sensor for the structural health monitoring. These "combo" stations will be used for the correction of nearby "single" stations of GNSS or strong-motion sensors (Psimoulis and Stiros, 2012) and finally the estimation of the displacement in full frequency range.

The combined use of GNSS and strong-motion sensors can contribute in the study of the soil-structure interaction during an earthquake by recording motions i) underground, below and around the structure foundations, ii) on the ground surface using it as reference level, iii) of indoor structures parts and iv) on top of structure, by using "combo" and "single" stations of GNSS and strong-motion sensors. For instance, for the case of monitoring of a tall building, "combo" GNSS and strong-motion stations will be set on the ground surface and on top of the structure. The "combo" stations will be used for the definition of the sensors consistency, for calibration of the one the other displacement amplitude and finally the estimation of the displacement of the corresponding monitoring points. The calibration of these GNSS and strong-motion records will be used for the corrections of the corresponding nearby GNSS/strong-motion "single" stations. The strong-motion "single" stations will be used as underground stations for soil motion and as intermediate floor indoor stations for the estimation of the displacement in the corresponding floors. The GNSS "single" stations can be used on the top of the building for the detection of potential rotations and tilts, which cannot be detected by the strong-motion sensors (Geng et al., 2014). The main advantage of this approach is the use of the consistency of the sensors for their calibration and correction and the estimation of the motion (3-D displacement and corresponding rotations) of the entire soil structure interaction system (from underground to top of the structure), limiting simultaneously the cost of the stations by using several "single" GNSS/strong-motion stations.

Apart from the structural health monitoring, significant contribution for the earthquake engineering would be the fast and reliable definition of displacement waveforms derived from combined after an earthquake. Furthermore, the comparison of the derived GNSS and strong-motion displacement waveforms, in combination with corresponding detailed geological maps may reveal local geological formations (basins, etc.), which amplify the seismic signal and will be valuable for the design of the structures.

## 7. Conclusions

We examined the consistency of GPS and strong-motion records for the recording of the megathrust earthquake of Mw9.0 Tohoku 2011, an extreme case for seismic rupture. We show that the frequency range of the consistent application of the two monitoring systems and how that depends on the motion frequency and the location relative to the excitation source. The two systems can supplement and correct successfully each other by using i) GPS for very low frequency (<0.01Hz) ii) combination of GPS and strong-motion/accelerometer for moderate frequencies (>0.01Hz and <0.3Hz) and iii) strong-motion/accelerometer for high frequencies (>0.3Hz), taken into consideration always the relative phase shift between the derived displacement of the two systems.

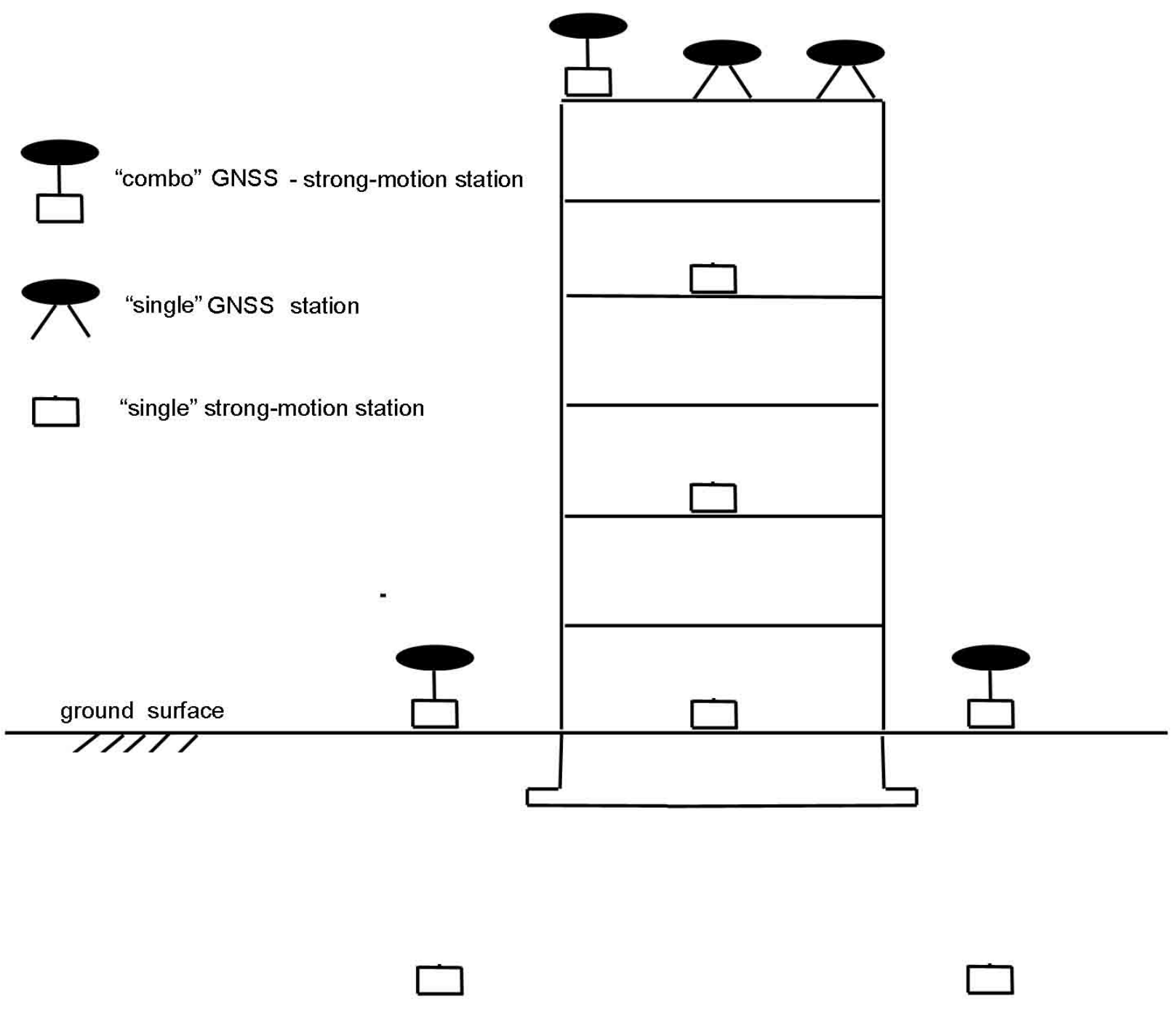

Fig. 8 Monitoring of the interaction of a tall building with soil during earthquake using stations of combinations of GNSS and strong-motion sensors

The conclusions of this study show that GPS and strong-motion/accelerometers can be used jointly supplementing each other by correcting one the other, providing more efficient structural health monitoring. However, the proper modeling of the phase shift/delay of the two systems is still very important task and will improve the fusion and merge of the two records (Chatzi and Smyth, 2009) even in real-time (Geng et al., 2013), affecting finally the efficiency of the structural health monitoring and the earthquake engineering.

## Acknowledgements

This study has been supported by Swiss National Fund grants in the framework of the “High-rate GNSS for Seismology” project.